\documentclass[twocolumn,prb,showpacs]{revtex4}
\usepackage{graphicx}
\usepackage{dcolumn}
\usepackage{bm}
\usepackage[cp1251]{inputenc}
\usepackage[russian]{babel}
\begin{document}

\title{The Theory of Light Scattering by Semiconductor Quantum
Dots.\\
Semiclassical Method Using Retarded Potentials}
\author {I. G. Lang, L. I. Korovin}
\address {A. F. Ioffe Physical-Technical Institute, Russian
Academy of Sciences, 194021 St. Petersburg, Russia}
\author {S. T. Pavlov\dag\ddag}
\address {\dag Facultad de Fisica de la UAZ, Apartado Postal C-580,
98060 Zacatecas, Zac., Mexico;\\ E-mail: pavlov@ahobon.reduaz.mx \\
\ddag P.N. Lebedev Physical Institute of Russian Academy of
Sciences. 119991, Moscow, Russia;\\ pavlov@sci.lebedev.ru}

\begin{abstract}
The theory of elastic light scattering by semiconductor quantum
dots is suggested. The semiclassical method, applying retarded
potentials to avoid the problem of bounder conditions for electric
and magnetic field, is used. The exact results for the Pointing
vector on large distances from a quantum dot, formulas of
differential cross sections of light scattering for the
monochromatic and pulse irradiation are obtained.
\end{abstract}

\pacs {78.47. + p, 78.66.-w}

\maketitle

\section{Introduction}

The excitonic excitations in size-quantized semiconductor objects
(quantum wells, quantum wires, quantum dots) may be investigated
by measuring of elastic light scattering by object mentioned
above. Light scattering increases resonantly when the stimulating
light frequency $\omega_\ell$ equals to the exciton frequency
 $\omega_0$. The width of the resonant peak is determined by the exciton damping $\Gamma$.

It was shown \cite{bb1} that  $\Gamma=\gamma_r+\gamma$, where
$\gamma_r (\gamma)$ is the radiative (nonradiative) exciton
damping. The same concept was extended on light absorption by
quantum wells \cite{bb2} (see also \cite{bb3}). Light reflection
by quantum wells, quantum wires, quantum dots was considered in
 \cite{bb41}.

There are two ways to investigate theoretically light scattering
by semiconductor quantum objects: quantum and semiclassical. The
quantum way applies the quantum perturbation theory. The electric
field is being quantized, processes of photon annihilation with an
exciton creation and vice versa are considered.  The quantum way
was used for calculations of a light scattering section by any
quantum dot. The quantum perturbation theory allows to obtain
results in the lowest order on the electron-photon interaction,
what excludes taking into account of the radiative damping
$\gamma_r$. The nonradiative damping $\gamma$ does not appear in
the quantum theory too, and the resonant denominator has a form
$(\omega_\ell-\omega_0)^2+\delta^2, \delta\rightarrow 0$, instead
of the precise expression $(\omega_\ell-\omega_0)^2+\Gamma^2/4$.
The advantage of the quantum theory is its simplicity.

With the help of the semiclassical method, classical electric and
magnetic fields are calculated, whereas the description of the
electron system remains quantum one (thus, there are essential the
nondiagonal matrix elements of the quasimomentum operator ${\bf
p}_{cv}$, corresponding to the electron-hole pair creation). The
semiclassical method starts since the calculation of the current
and charge densities averaged on the crystal ground state
 \footnote{The Raman scattering is determined by fluctuations
  of current and charge densities.}
  and induced by the electric field inside of the semiconductor
object \cite{bb6,bb7}. When the stimulating electric field ${\bf
E}_0({\bf r}, t)$ is used in expressions for these densities, we
are limited again by the lowest order on the electron-light
interaction.  All the orders on the electron-light interaction are
taken into account, when the genuine electric field is substituted
into these densities.

Further, two versions of the semiclassical method may be applied.
The first version is a solution of the Maxwell equations inside
and outside of an object and application of the bounder conditions
for electric and magnetic fields on borders of the object. This
version is convenient when the refraction coefficients inside
$\nu$ and outside $\nu_0$ of the object are different. In such a
way the problem of light reflection and absorption by wide quantum
wells (the width of which may be comparabel with the light wave
length) \cite{bb8,bb9,bb10}. The electric fields arising in the
resonant light scattering by an exciton in a spherical quantum dot
(consisting of a cubical crystal of $T_d$ class (for example,
GaAs) and limited by the infinitely rectangular potential barrier
under condition $\nu_0\neq\nu$) are calculated \cite{bb11} by
similar way. However, in the case of spherical quantum dots,
application of the boundary conditions results into extremely
bulky calculations, and in cases of quantum dots of other forms
become even more complicated.

Therefore, we suggest a second version of the semiclassical method
- the method of the retarded potential. The essence of the method
is as follows. We substitute the values of the averaged induced
densities of currents and charges into expressions for the vector
and scalar potentials (see (7) and (8) below). Having used
potentials, containing the genuine electric field inside  and
electric and magnetic field outside of the object, we obtain an
integral equation for the electric field inside of the object,
which may be solved in some cases. Further, we determine the
fields outside of the object. The advantage of the method of
retarded potentials is a possibility to avoid the problem of the
boundary conditions.

For the case of a normal incidence of light at the quantum well
surface, the method is described in \cite{bb7} and applied in
\cite{bb12}. It was shown that the both versions of the
semiclassical method lead to identical results. Calculations
similar to the method of the retarded potentials are used in
\cite{bb4}.

Below, we consider the light scattering by a quantum dot of
arbitrary form. We choose the electric field ${\bf E}_0({\bf r},
t)$ in the form, allowing to obtain results for monochromatic and
pulse irradiation.

\section{Current and charge densities inside of a quantum dot Плотности тока и заряда внутри КТ}

According to \cite{bb7} (see Appendix 1), the average density of
induced current inside of a quantum dot equals
\begin{eqnarray}
\label{l}&&{\bf j}({\bf r}, t)=\langle 0|{\bf j}_1({\bf r},
t)|0\rangle={ie^2\over 2\pi\hbar\omega_gm_0^2}\sum_\eta
F_\eta({\bf r})\nonumber\\
&&\times\int d^3{\bf r}^\prime\int_{-\infty}^\infty
dt^\prime\int_{-\infty}^\infty d\omega e^{-i\omega(t-t^\prime)}\nonumber\\
&&\times\left\{{{\bf p}_{cv\eta}^*({\bf p}_{cv\eta}{\bf E}_0({\bf
r}^\prime, t^\prime))\over\omega-\omega_\eta+i\gamma_\eta/2}
+{{\bf p}_{cv\eta}({\bf p}^*_{cv\eta}{\bf E}_0({\bf r}^\prime,
t^\prime))\over\omega-\omega_\eta+i\gamma_\eta/2}\right\},~~~~
\end{eqnarray}
where $e(m_0)$ is the free electron charge (mass), $\hbar\omega_g$
is the energy gap, $\eta$ is the exciton index, $F_\eta({\bf r})$
is the real ("envelope") exciton wave function at ${\bf r}_e={\bf
r}_h={\bf r}, {\bf r}_e ({\bf r}_h)$ is the electron (hole)
radius-vector, ${\bf p}_{cv\eta}$ is the interband matrix element
of the quasi-momentum operator, ${\bf E}({\bf r}, t)$ is the
electric field inside of a quantum dot; $\hbar\omega_\eta,
\gamma_\eta$ is the energy (counted from the ground state energy)
and exciton nonradiative damping, respectively.

Let us introduce the Fourier-transform ${\bf {\cal E}}({\bf r},
\omega)$ of the electric field \footnote{A fragmentation of
contributions  in (2) is made so that in the zero approximation
(see (31) below) we have ${\bf {\cal E}}({\bf r}, \omega)=2\pi
{\bf e}_\ell E_0e^{i\kappa z D_0(\omega)}$.}
\begin{equation}
\label{2}{\bf E}({\bf r}, t)={1\over 2\pi}\int_{-\infty}^\infty
d\omega e^{-i\omega t}{\bf {\cal E}}({\bf r}, \omega)+c.c. .
\end{equation}
Then, we have from (1)
\begin{eqnarray}
\label{3}&&{\bf j}({\bf r}, t)={ie^2\over
2\pi\hbar\omega_gm_0^2}\int_{-\infty}^\infty d\omega\sum_\eta
F_\eta({\bf r})\nonumber\\
&&\times \left\{{{\bf
p}_{cv\eta}^*M_\eta\over\omega-\omega_\eta+i\gamma_\eta/2} +{{\bf
p}_{cv\eta}\tilde{M}_\eta\over\omega+\omega_\eta+i\gamma_\eta/2}
\right\},
\end{eqnarray}
where
\begin{eqnarray}
\label{4}M_\eta=\int d^3rF_\eta({\bf r})({\bf p}_{cv\eta}{\bf {\cal E}}({\bf r}, \omega)),\nonumber\\
{\tilde{M}}_\eta=\int d^3rF_\eta({\bf r})({\bf p}^*_{cv\eta}{\bf
{\cal E}}({\bf r}, \omega)).
\end{eqnarray}
We determine the average density of induced charge inside of a
quantum dot from the continuity equation
\begin{equation}
\label{5}div~ {\bf j}({\bf r}, t)+{\partial{\bf \rho}({\bf r},
t)\over\partial t}=0
\end{equation}
as
\begin{eqnarray}
\label{6}&&{\bf \rho}({\bf r}, t)={e^2\over
2\pi\hbar\omega_gm_0^2}\int_{-\infty}^\infty
{d\omega\over\omega}\nonumber\\
&&\times\sum_\eta\left\{div~ ({\bf p}_{cv\eta}^*F_\eta({\bf r}))
{M_\eta\over
\omega-\omega_\eta+i\gamma_\eta/2}\right.\nonumber\\
 &&\left.+div~ ({\bf
p}_{cv\eta} F_\eta({\bf r})){{\tilde M}_\eta \over
\omega+\omega_\eta+i\gamma_\eta/2}\right\}+c.c.
\end{eqnarray}
The genuine electric field inside of a quantum dot appears in the
RHSs of expressions (1),(3) and (6). We assume that functions
$F({\bf r})$ and $dF({\bf r})/d{bf r}$ approach to 0 at
$r\rightarrow\infty$, whence it follows that the current and
charge densities equal 0 at $r\rightarrow\infty$.

\section{Retarded potentials}

Vector and scalar potentials, corresponding to the induced
electric field inside and outside of a quantum dot are defined by
formulas \cite{bb13}
\begin{equation}
\label{7} {\bf A}({\bf r}, t)={1\over c}\int d^3r^\prime{{\bf
j}({\bf r}^\prime, t-\nu|{\bf r}-{\bf r}^\prime|/c)\over|{\bf
r}-{\bf r}^\prime|},
\end{equation}
\begin{equation}
\label{8}\varphi({\bf r}, t)={1\over\nu^2}\int
d^3r^\prime{\rho({\bf r}^\prime, t-\nu|{\bf r}-{\bf
r}^\prime|/c)\over|{\bf r}-{\bf r}^\prime|},
\end{equation}
where $\nu$ is the refraction coefficient, which is identical
inside and outside of a quantum dot. Substituting (3) and (6) in
(7) and (8), we obtain
\begin{eqnarray}
\label{9}&&{\bf A}({\bf r}, t)={ie^2\over
2\pi\hbar\omega_gm_0^2c}\int_{-\infty}^\infty d\omega e^{-i\omega
t}\nonumber\\
&&\times\sum_\eta\left\{{{\bf
p}_{cv\eta}^*M_\eta\over\omega-\omega_\eta+i\gamma_\eta/2}+ {{\bf
p}_{cv\eta}{\tilde M}_\eta\over\omega+\omega_\eta+i\gamma_\eta/2}
\right\}\nonumber\\
&&\times\int d^3r^\prime F_\eta({\bf r}^\prime){e^{i\kappa|{\bf
r}-{\bf r}^\prime|}\over|{\bf r}-{\bf r}^\prime|}+c.c.,
\end{eqnarray}
\begin{eqnarray}
\label{10}&&\varphi({\bf r}, t)={ie^2\over
2\pi\hbar\omega_gm_0^2\nu^2}\int_{-\infty}^\infty
{d\omega\over\omega} e^{-i\omega
t}\nonumber\\
&&\times\sum_\eta\left\{{{\bf
p}_{cv\eta}^*M_\eta\over\omega-\omega_\eta+i\gamma_\eta/2}+ {{\bf
p}_{cv\eta}{\tilde M}_\eta\over\omega+\omega_\eta+i\gamma_\eta/2}
\right\}\nonumber\\
&&\times\int d^3r^\prime {e^{i\kappa|{\bf r}-{\bf
r}^\prime|}\over|{\bf r}-{\bf r}^\prime|}{d\over d{\bf
r}^\prime}F_\eta({\bf r}^\prime)+c.c.,
\end{eqnarray}
where $\kappa=\omega\nu/c$. In the RHS of (10) we transfer the
derivative  on ${\bf r}^\prime$ with the opposite sign at the
factor $e^{i\kappa|{\bf r}-{\bf r}^\prime|}/|{\bf r}-{\bf
r}^\prime|$, and apply the ratio
$${d\over d{\bf
r}^\prime}{e^{i\kappa|{\bf r}-{\bf r}^\prime|}\over|{\bf r}-{\bf
r}^\prime|}=-{d\over d{\bf r}}{e^{i\kappa|{\bf r}-{\bf
r}^\prime|}\over|{\bf r}-{\bf r}^\prime|}.$$ Получаем
\begin{eqnarray}
\label{11}&&\varphi({\bf r}, t)={ie^2\over
2\pi\hbar\omega_gm_0^2\nu^2}\int_{-\infty}^\infty
{d\omega\over\omega} e^{-i\omega
t}\nonumber\\
&&\times\sum_\eta\left\{{{\bf
p}_{cv\eta}^*M_\eta\over\omega-\omega_\eta+i\gamma_\eta/2}+ {{\bf
p}_{cv\eta}{\tilde M}_\eta\over\omega+\omega_\eta+i\gamma_\eta/2}
\right\}\nonumber\\
&&\times{d\over d{\bf r}}\int d^3r^\prime {e^{i\kappa|{\bf r}-{\bf
r}^\prime|}\over|{\bf r}-{\bf r}^\prime|}F_\eta({\bf
r}^\prime)+c.c..
\end{eqnarray}
\section{Electric and magnetic fields}

Electric and magnetic fields are defined as
\begin{eqnarray}
\label{12}{\bf E}({\bf r}, t)&=&{1\over c}{\partial{\bf E}({\bf
r}, t)\over\partial t}-{\partial\varphi({\bf r}, t)\over\partial
{\bf
r}},\nonumber\\
{\bf H}({\bf r}, t)&=&rot {\bf A}.
\end{eqnarray}
Let us calculate the fields ${\bf E}({\bf r}, t)$ and ${\bf
H}({\bf r}, t)$   at very large distances  $r$ from a quantum dot,
exceeding all the values of the length dimensionality, essential
in our problem. It means that we determine fields in the limit
$r\rightarrow\infty$.

Let us calculate the integral
\begin{equation}
\label{13}J_\eta({\bf r})=\int d^3r^\prime F_\eta({\bf r}^\prime)
{e^{i\kappa|{\bf r}-{\bf r}^\prime|}\over|{\bf r}-{\bf
r}^\prime|}.
\end{equation}
Since in the limit $r\rightarrow\infty$
$$e^{i\kappa|{\bf r}-{\bf r}^\prime|}\simeq e^{i\kappa r - \kappa|{\bf r}-{\bf
r}^\prime|/r}$$, we obtain the result
\begin{equation}
\label{14}J_\eta({\bf r})|_{r\rightarrow\infty}={e^{i\kappa
r}\over r} P_\eta({\bf \kappa}_s),
\end{equation}
where the designations
\begin{eqnarray}
\label{15}&&{\bf \kappa}_s=\kappa{\bf r}/r,\nonumber\\
&& P_\eta({\bf \kappa})=\int d^3r F_\eta({\bf r}) e^{-i{\bf
\kappa}{\bf r}},
\end{eqnarray}
are introduced, and ${\bf \kappa}_s$ is the wave vector of the
scattered light. We obtain also
\begin{equation}
\label{16}\left({\partial\over\partial r}J_\eta({\bf
r})\right)_{r\rightarrow\infty}=i\kappa{{\bf r}\over r}{e^{i\kappa
r}\over r} P_\eta({\bf \kappa}_s),
\end{equation}
and so on.

Using (14) and (16) and omitting contributions, small at
 $r\rightarrow\infty$, we obtain for the fields:
\begin{eqnarray}
\label{17}&&{\bf E}({\bf r}, t)={e^2\over
2\pi\hbar\omega_gm_0^2c^2}\int_{-\infty}^\infty d\omega\omega
e^{-i\omega
t}{e^{i\kappa r}\over r}\sum_\eta P_\eta({\bf \kappa}_s)\nonumber\\
&&\times\left\{ {[({\bf p}_{cv}^*{{\bf r}\over r}){{\bf r}\over
r}-{\bf p}_{cv}^*] M_\eta\over\omega-\omega_\eta+i\gamma_\eta/2}
+{[({\bf p}_{cv}{{\bf r}\over r}){{\bf r}\over r}-{\bf p}_{cv}]
{\tilde M}_\eta\over\omega+\omega_\eta+i\gamma_\eta/2}\right\}\nonumber\\
&&+c.c.,
\end{eqnarray}
\begin{eqnarray}
\label{18}&&{\bf H}({\bf r}, t)={e^2\nu\over
2\pi\hbar\omega_gm_0^2c^2}\int_{-\infty}^\infty d\omega\omega
e^{-i\omega
t}{e^{i\kappa r}\over r}\sum_\eta P_\eta({\bf \kappa}_s)\nonumber\\
&&\times\left\{ {[{\bf p}_{cv}^*\times{{\bf r}\over r}]
M_\eta\over\omega-\omega_\eta+i\gamma_\eta/2} +{[{\bf
p}_{cv}\times{{\bf r}\over r}] {\tilde
M}_\eta\over\omega+\omega_\eta+i\gamma_\eta/2}\right\} +c.c..
\end{eqnarray}
It follows from (17) and (18) that at $r\rightarrow\infty$ the
fields ${\bf E}({\bf r}, t)$ and ${\bf H}({\bf r}, t)$ are
perpendicular to the radius-vector ${\bf r}$.

\section{Decomposition of electric and magnetic fields on the components
with left and right circular polarizations}

Let us introduce the circular polarization vectors
$${\bf e}^\pm_s={1\over\sqrt 2}({\bf e}_x\pm i{\bf e}_y)$$,
where the unite vectors  ${\bf e}_x$ and ${\bf e}_y$ are
perpendicular to the axis $z$, aligned parallel to ${\bf r}$.

Let us write down the fields as
\begin{eqnarray}
\label{19}{\bf E}({\bf r}, t)_{r\rightarrow\infty}&=&{\bf
E}_c({\bf r}, t)+{\bf E}^*_c({\bf
r}, t),\nonumber\\
{\bf H}({\bf r}, t)_{r\rightarrow\infty}&=&{\bf H}_c({\bf r},
t)+{\bf H}^*_c({\bf r}, t),
\end{eqnarray}
and decompose them on polarizations
\begin{eqnarray}
\label{20}{\bf E}_c({\bf r}, t)&=&E^+({\bf r}, t){\bf
e}^+_s+E^-({\bf r}, t){\bf e}^-_s,\nonumber\\
{\bf H}_c({\bf r}, t)&=&H^+({\bf r}, t){\bf e}^+_s+H^-({\bf
r},t){\bf e}^-_s.
\end{eqnarray}
Multiplying both parts of (20) consecutively on ${\bf e}^+_s$ and
${\bf e}^{-}_s$, we find
\begin{eqnarray}
\label{21}{\bf E}^\pm({\bf r}, t)&=&-{e^2\over
2\pi\hbar\omega_gm_0^2c^2r}\int_{-\infty}^\infty d\omega\omega
e^{i\kappa r-i\omega
t}\nonumber\\
&&\times Q({\bf \kappa}_s,\omega, {\bf e}^\pm_s),
\end{eqnarray}
\begin{equation}
\label{22}H^\pm({\bf r}, t)=\mp i\nu E^\pm ({\bf r}, t),
\end{equation}
where
\begin{eqnarray}
\label{23} &&Q({\bf \kappa}_s,\omega, {\bf e}^\pm_s)=\sum_\eta P_\eta({\bf \kappa}_s)\nonumber\\
&&\times\left\{ {({\bf p}_{cv\eta}{\bf e}_s)^*
M_\eta\over\omega-\omega_\eta+i\gamma_\eta/2} +{({\bf
p}_{cv\eta}{\bf e}^{\pm *}){\tilde
M}_\eta\over\omega+\omega_\eta+i\gamma_\eta/2}\right\} +c.c..~~~~
\end{eqnarray}
\section{The Pointing vector}

 The Pointing vector is equal \cite{bb13}
\begin{equation}
\label{24}{\bf S}({\bf r}, t)={c\over 4\pi}{\bf E}({\bf r},
t)\times {\bf H}({\bf r}, t).
\end{equation}
Substituting (19) in (24), we obtain
\begin{eqnarray}
\label{25}&&{\bf S}({\bf r}, t)_{r\rightarrow\infty}={c\over
4\pi}\nonumber\\&&\times\{{\bf E}_c({\bf r}, t)\times {\bf
H}_c^*({\bf r}, t)+{\bf E}_c^*({\bf r}, t)\times {\bf H}_c({\bf
r}, t)\nonumber\\&&+{\bf E}_c({\bf r}, t)\times {\bf H}_c({\bf r},
t)+{\bf E}_c^*({\bf r}, t)\times {\bf H}_c^*({\bf r}, t)\}.
\end{eqnarray}
Two first terms in the RHS of(25) provide the main contribution
into the Pointing vector, averaged on time. \footnote{We assume
averaging on comparatively short time, exceeding the period
$T_\ell=2\pi/\omega_\ell$, where $\omega_\ell$ is the light
frequency of a monochromatic irradiation or carrying frequency of
a pulse irradiation.} Restricting by that contribution, we have
\begin{equation}
\label{26}{\bf S}({\bf r}, t)_{r\rightarrow\infty}=\sum_\mu {\bf
S}({\bf r}, t, {\bf l}_s),
\end{equation}
where $\sum_\mu$  designates  summation on circular polarizations
of scattered light,
\begin{eqnarray}
\label{27}&&{\bf S}({\bf r}, t, {\bf
l}_s)_{r\rightarrow\infty}={c\over (2\pi)^3}\left({e^2\over\hbar
c} \right)^2{\nu\over m_gm_0^4c}{{\bf r}\over r}\nonumber\\
&&\times\left|\int^\infty_{-\infty}d\omega \omega e^{i\kappa
r-i\omega t}Q({\bf \kappa}_s, \omega, {\bf e}_s)\right|^2.
\end{eqnarray}
Opening the square of module in the RHS of (27), we obtain
\begin{eqnarray}
\label{28}\int_{-\infty}^\infty d\omega\int_{-\infty}^\infty
d\omega^\prime \omega\omega^\prime e^{i(\kappa-\kappa^\prime)
r-i(\omega-\omega^\prime) t}\nonumber\\\times Q({\bf \kappa}_s,
\omega, {\bf e}_s)Q^*({\bf \kappa}_s^\prime, \omega^\prime, {\bf
e}_s),
\end{eqnarray}
where $\kappa^\prime=\omega^\prime\nu/c, {\bf
\kappa}_s^\prime=\kappa_s^\prime{\bf r}/r$. As it will be seen
below, the value $Q({\bf \kappa}_s, \omega, {\bf e}_s)$ always
contains the factor $D_0(\omega)$, describing the form of the
stimulating pulse ${\bf E}_0({\bf r}, t)$, defined as
\begin{equation}
\label{29}{\bf E}_0({\bf r}, t)=E_0{\bf
e}_\ell\int^\infty_{-\infty}d\omega e^{i{\bf \kappa}_\ell{\bf
r}-i\omega t}D_0(\omega)+c.c.,
\end{equation}
where ${\bf \kappa}_\ell$ is the vector, which magnitude is equal
 $\omega_\ell\nu/c$, and the direction is defined by the direction of stimulating
 light.
For the monochromatic irradiation
\begin{equation}
\label{30}D_0(\omega)=\delta(\omega-\omega_\ell),
\end{equation}
and for the pulse irradiation the frequency is spreaded beside
$\omega_\ell$  for some interval $\Delta\omega$, inversely
proportional to the pulse duration $\Delta t$. In any case, the
carrying frequency $\omega_\ell$ disappears from (28) due to the
factor $e^{-i(\omega-\omega^\prime)t}$ in the integrand. The
contribution of the last two terms from the RHS of (25) contains
under integral $e^{-i(\omega+\omega^\prime)t}$ , what makes this
contribution rapidly oscillating in time and approaching 0 at
averaging on time.

\section{The scattering cross section}

In a case of monochromatic irradiation, when condition (30) is
satisfied, it is conveniently to introduce a concept of the cross
section of light scattering on a quantum dot.

The differential cross section is defined as a ratio of a
magnitude of scattered energy flux into an solid angle interval
$do_s$ to the magnitude of incident energy flux on the area unite
in a time unite. For a monochromatic irradiation, it is equal
\begin{equation}
\label{31}{\bf S}_0={c\nu\over2\pi}E_0^2.
\end{equation}
Using (26) and (27), we obtain
\begin{eqnarray}
\label{32}&&{d\sigma\over do_s}=\sum_\mu{d\sigma_\mu\over
do_s},\nonumber\\
&&{d\sigma_\mu\over do_s}={1\over (2\pi)^2}\left({e^2\over\hbar c}
\right)^2{1\over
\omega_g^2m_0^4c^2E_0^2}\nonumber\\
&&\times\left|\int_{-\infty}^\infty d\omega\omega e^{i(\kappa
r-i\omega t}Q({\bf \kappa}_s, \omega, {\bf e}_s) \right|^2.
\end{eqnarray}
In a case of monochromatic irradiation, the factor
$\delta(\omega-\omega_\ell)$ presents always in the expression for
 $Q({\bf
\kappa}_s, \omega^\prime), {\bf e}_s) $ (see below). It means that
we can write
\begin{equation}
\label{33}Q({\bf \kappa}_s, \omega, {\bf e}_s)=q({\bf \kappa}_s,
\omega_\ell, {\bf e}_s)\delta(\omega-\omega_\ell),
\end{equation}
and then (32) is transformed to the form
\begin{equation}
\label{34}{d\sigma_\mu\over do_s}={1\over
(2\pi)^2}\left({e^2\over\hbar c} \right)^2{\omega^2_\ell\over
\omega_g^2m_0^4c^2E_0^2} \left|q({\bf \kappa}_s, \omega_\ell, {\bf
e}_s) \right|^2,
\end{equation}
and dependencies on $r$ and $t$ disappear.

\section{The lowest approximation on the electron-light interaction}

In the lowest approximation the genuine electric field inside of a
quantum dot, which is included in expressions for $M_\eta(\omega)$
and ${\tilde M}_\eta(\omega)$, is substituted by the stimulating
field. According to (29) and (2), the Fourier transform is as
follows
\begin{equation}
\label{35}{\bf {\cal E}}_0({\bf r}, \omega,)=2\pi {\bf e}_\ell
E_0e^{i{\bf \kappa_\ell}{\bf r}}D_0(\omega).
\end{equation}
Then,
\begin{eqnarray}
\label{36}&&M_{\eta 0}=2\pi E_0D_0(\omega) P_\eta^*({\bf
\kappa_\ell})({\bf e}_\ell{\bf p}_{cv\eta}),\nonumber\\
&&{\tilde {M}}_{\eta 0}=2\pi E_0D_0(\omega) P_\eta^*({\bf
\kappa_\ell})({\bf e}_\ell{\bf p}_{cv\eta}^*),\nonumber\\
\end{eqnarray}
\begin{eqnarray}
\label{37} &&Q_0=2\pi E_0D_0(\omega) \sum_\eta P_\eta({\bf
\kappa}_s) P_\eta^*({\bf
\kappa}_\ell)\nonumber\\
&&\times\left\{{({\bf e}_s{\bf p}_{cv\eta}^*)({\bf e}_\ell{\bf
p}_{cv\eta})\over \omega-\omega_\eta+i\gamma_\eta/2} + {({\bf
e}_s^*{\bf p}_{cv\eta})({\bf e}_\ell{\bf p}^*_{cv\eta})\over
\omega+\omega_\eta+i\gamma_\eta/2}\right\}.
\end{eqnarray}
Substituting (37) in (21), (22) and (27), we obtain electric and
magnetic fields and the Pointing vector in the limit
$r\rightarrow\infty$.

For the monochromatic irradiation, we obtain the differential
section
\begin{equation}
\label{38}{d\sigma_{\mu 0}\over do_s}=\left({e^2\over\hbar c}
\right)^2{\omega^2_\ell\over \omega_g^2m_0^4c^2} \left|\sum_\eta
P_\eta({\bf {\cal \kappa}}_s) P^*_\eta({\bf {\cal
\kappa}}_\ell)({\bf v}_\eta{\bf e}_s^*) \right|^2,
\end{equation}
where
\begin{eqnarray}
\label{39} &&|{\bf {\cal \kappa}}_s|=|{\bf {\cal \kappa}}_\ell|={\omega_\ell\nu\over c},\nonumber\\
&&{\bf v}_\eta={{\bf p}^*_{cv\eta}({\bf e}_\ell{\bf
p}_{cv\eta})\over \omega-\omega_\eta+i\gamma_\eta/2} + {{\bf
p}_{cv\eta}({\bf e}_\ell{\bf p}^*_{cv\eta})\over
\omega+\omega_\eta+i\gamma_\eta/2}.
\end{eqnarray}
\section{Comparison of results of semiclassical and quantum  theories \cite{bb14}}

Let us omit in (38) the nonresonant term, containing
$(\omega+\omega_\eta+i\gamma_\eta/2)^{-1}$. We obtain
\begin{eqnarray}
\label{40}&&{d\sigma_{\mu 0}\over do_s}=\left({e^2\over\hbar c}
\right)^2{\omega^2_\ell\over \omega_g^2m_0^4c^2}\nonumber\\
&&\times \left|\sum_\eta {P_\eta({\bf {\cal \kappa}}_s)
P^*_\eta({\bf {\cal \kappa}}_\ell)({\bf p}_{cv\eta}{\bf
e}_\ell)({\bf p}_{cv\eta}{\bf
e}_s)^*\over\omega-\omega_\eta+i\gamma_\eta/2} \right|^2.
\end{eqnarray}
It follows from (40) that if the light wave length exceeds the
quantum dot sizes and
$$P_\eta({\bf {\cal \kappa}}_\ell)\simeq P_\eta({\bf {\cal \kappa}}_s)\simeq P_\eta(0),$$
the polarization and angular distribution of scattered light
depend only on vectors  ${\bf p}_{cv\eta}$, corresponding to the
exciton, which is in resonance with the stimulating light. This
result is very close to results of the quantum theory  (see
 (15) from \cite{bb15}), but the factor $(\omega_\ell/\omega_g)^4$
is substituted by $(\omega_\ell/\omega_g)^2$ , what is
incidentally under resonant conditions, and the value
$(\omega-\omega_\eta+i\delta)^{-1}$ is substituted by
$(\omega-\omega_\eta+i\gamma_\eta/2)^{-1}$. Obviously, the
semiclassical method provides more exact results even in the
lowest approximation on the electron-light interaction, since it
allows to include into the theory the nonradiative damping
 $\gamma_\eta$ (which determines light absorption by a quantum dot). Besides,
 Кроме того, the semiclassical method allows to investigate light scattering and absorption
for a pulse irradiation.

Further, we pass on to the precise semiclassical theory, taking
into account all the orders on the electron-light interaction.

\section{The equation for electric field inside of a quantum dot}

First of all, we calculate defined in (4)  $M_\eta$ and ${\tilde
M}_\eta$ from the RHS of (23), i.e., we substitute in (4) the
Fourier-components of the genuine electric field inside of the
quantum dot. We obtain an equation for these Fourier-components
 ${\bf {\cal E}}({\bf r}, \omega)$ as follows, writing the genuine electric field inside of the
quantum dot in the form
\begin{equation}
\label{41} {\bf E}({\bf r}, t)={\bf E}_0({\bf r}, t)-{1\over
c}{\partial{\bf A}({\bf r}, t)\over\partial
t}-{\partial\varphi({\bf r}, t)\over\partial {\bf r}},
\end{equation}
where ${\bf E}_0({\bf r}, t)$ is the stimulating electric field
(29), ${\bf A}({\bf r}, t)$ and  $\varphi({\bf r}, t)$ are the
vector and scalar potentials, defined in (9) and (10) and
corresponding to the secondary radiation of the quantum dot. For
the Fourier-components ${\bf {\cal E}}({\bf r}, \omega)$   of the
field (41), we obtain the integral equation
\begin{eqnarray}
\label{42}&&{\bf {\cal E}}({\bf r}, \omega)=2\pi{\bf e}_\ell
E_0e^{i{\bf \kappa}_\ell{\bf r}}D_0(\omega) -{e^2\over\hbar c}
{\omega\over
\omega_gm_0^2c}\nonumber\\
&&\times\sum_\eta\left\{{M_\eta\left[{\bf
p}_{cv\eta}^*+\kappa^{-2}{\partial\over\partial{\bf r}}
\left({\partial{\bf p}_{cv\eta}^*\over\partial{\bf
r}}\right)\right]\over\omega-\omega_\eta+i\gamma_\eta/2}\right.\nonumber\\
&&+\left.{{\tilde M}_\eta\left[{\bf
p}_{cv\eta}+\kappa^{-2}{\partial\over\partial{\bf r}}
\left({\partial{\bf p}_{cv\eta}\over\partial{\bf
r}}\right)\right]\over\omega+\omega_\eta+i\gamma_\eta/2}\right\}\Phi_\eta({\bf
r}),
\end{eqnarray}
where
\begin{equation}
\label{43} \Phi_\eta({\bf r})=\int d^3rF_\eta({\bf
r}^\prime){e^{i\kappa|{\bf r}-{\bf r}^\prime|}\over|{\bf r}-{\bf
r}^\prime|}.
\end{equation}

\section{The equation system for coefficients $M_\eta$ and ${\tilde M}_\eta$}

We multiply (42) at $F_{\eta^\prime}({\bf r}){\bf
p}_{cv\eta^\prime}$  and integrate on ${\bf r}$, then, perform the
same operation, using $F_{\eta^\prime}({\bf r}){\bf
p}^*_{cv\eta^\prime}$. We obtain the following system of equations
for $M_\eta$ and ${\tilde M}_\eta$ \footnote{Tilde over a letter
means, as in (4), the replacement of vectors  ${\bf p}_{cv\eta}$
by ${\bf p}_{cv\eta}^*$ and vice versa. }
\begin{eqnarray}
\label{44}&&M_{\eta^\prime}=2\pi E_0D_0(\omega)({\bf e}_\ell{\bf
p}_{cv\eta^\prime})P_{\eta^\prime}^*(\kappa_\ell)-{e^2\over\hbar c}{\omega\over \omega_gm_0^2c}\nonumber\\
&&\times \sum_\eta\left \{{M_\eta
R_{\eta\eta^\prime}\over\omega-\omega_\eta+i\gamma_\eta/2
}+{{\tilde M}_\eta
S_{\eta\eta^\prime}\over\omega+\omega_\eta+i\gamma_\eta/2 }\right
\},\nonumber\\
&&{\tilde M}_{\eta^\prime}=2\pi E_0D_0(\omega)({\bf e}_\ell{\bf
p}^*_{cv\eta^\prime})P_{\eta^\prime}^*(\kappa_\ell)-{e^2\over\hbar c}{\omega\over \omega_gm_0^2c}\nonumber\\
&&\times \sum_\eta\left \{{M_\eta {\tilde
S}_{\eta\eta^\prime}\over\omega-\omega_\eta+i\gamma_\eta/2
}+{{\tilde M}_\eta {\tilde
R}_{\eta\eta^\prime}\over\omega+\omega_\eta+i\gamma_\eta/2 }\right
\},
\end{eqnarray}
where
\begin{eqnarray}
\label{45}&&R_{\eta\eta^\prime}=({\bf p}^*_{cv\eta}{\bf
p}_{cv\eta^\prime})\int d^3rF_{\eta^\prime}({\bf r})\Phi_\eta({\bf
r})\nonumber\\
&&+{1\over\kappa^2}\int d^3rF_{\eta^\prime}({\bf
r})\left({\partial{\bf p}_{cv\eta^\prime}\over\partial{\bf
r}}\right)\left({\partial{\bf p}^*_{cv\eta}\over\partial{\bf
r}}\right)\Phi_\eta({\bf
r}), \nonumber\\
&&S_{\eta\eta^\prime}=({\bf p}_{cv\eta}{\bf
p}_{cv\eta^\prime})\int d^3rF_{\eta^\prime}({\bf r})\Phi_\eta({\bf
r})\nonumber\\
&&+{1\over\kappa^2}\int d^3rF_{\eta^\prime}({\bf
r})\left({\partial{\bf p}_{cv\eta^\prime}\over\partial{\bf
r}}\right)\left({\partial{\bf p}_{cv\eta}\over\partial{\bf
r}}\right)\Phi_\eta({\bf r}).
\end{eqnarray}
The integrals on ${\bf r}$  from (45) may be written in a more
symmetrical form
\begin{eqnarray}
\label{46}&&\int d^3rF_{\eta^\prime}({\bf r}) \Phi_\eta({\bf
r})=\int d^3r \int d^3r^\prime F_{\eta^\prime}({\bf
r})F_{\eta}({\bf r})\nonumber\\
&&\times{e^{i\kappa|{\bf r}-{\bf r}^\prime|}\over|{\bf
r}-{\bf r}^\prime|},\nonumber\\
&&\int d^3r F_{\eta^\prime}({\bf r})\left({\partial{\bf
p}_{cv\eta}\over\partial{\bf r}}\right)\left({\partial{\bf
p}^*_{cv\eta}\over\partial{\bf r}}\right)\Phi_\eta({\bf
r})\nonumber\\
&&=\int d^3r \int d^3r^\prime F_{\eta^\prime}({\bf
r})F_{\eta}({\bf r})\left({\bf
p}_{cv\eta^\prime}{\partial\over\partial{\bf r}}\right)\left({\bf
p}_{cv\eta^\prime}{\partial\over\partial{\bf
r}}\right)\nonumber\\
&&\times{e^{i\kappa|{\bf r}-{\bf r}^\prime|}\over|{\bf r}-{\bf
r}^\prime|}.
\end{eqnarray}
The coefficients from (44) have properties
\begin{equation}
\label{47} R_{\eta^\prime\eta}={\tilde R}_{\eta\eta^\prime},
S_{\eta^\prime\eta}=S_{\eta\eta^\prime}.
\end{equation}
\section{Calculation of the function $\Phi({\bf r})$}

Let us obtain the Fourier-transformation of the function
$$\Phi({\bf r})=\int d^3r^\prime F({\bf
r}^\prime) {e^{i\kappa |{\bf r}-{\bf r}^\prime|}\over |{\bf
r}-{\bf r}^\prime|}.$$ We obtain
\begin{equation}
\label{48}\Phi({\bf r})={1\over (2\pi)^3}\int d^3\kappa F({\bf
\kappa})f_\kappa({\bf \kappa})e^{i{\bf \kappa}{\bf r}},
\end{equation}
where
$$F_\kappa({\bf \kappa})=\int d^3r e^{-i{\bf \kappa}r}F({\bf
r})=P({\bf \kappa}), f_\kappa({\bf \kappa})$$
$$=\int d^3r
e^{-i{\bf \kappa}r}{e^{i\kappa r}\over r}.$$ Integration on ${\bf
r}$ results in
\begin{eqnarray}
\label{49}&&f_\kappa({\bf k})=f_\kappa(k)\nonumber\\
&&={2\pi\over k}\left\{{P\over k-\kappa}+ {P\over
k+\kappa}+i\pi\delta(k-\kappa)-{P\over
k+\kappa}\right\},\nonumber\\
&&{P\over k-\kappa}={1\over 2}\left({1\over
k-\kappa+i\delta}+{P\over k-\kappa-i\delta}\right),\nonumber\\
&&\delta\rightarrow +0.
\end{eqnarray}
The last term in the RHS of (49) may be omitted, then
\begin{eqnarray}
\label{50}\Phi({\bf r})={1\over (2\pi)^3}\int d^3k e^{i{\bf k}{\bf
r}}P({\bf k})\nonumber\\
\times\left\{4\pi{P\over
k^2-\kappa^2}+{2i\pi^2\over\kappa}\delta(k-\kappa) \right\}.
\end{eqnarray}
It is easy to check that the function $\Phi({\bf r})$ satisfies
the equation
\begin{equation}
\label{51}(\kappa^2+\Delta)\Phi({\bf r})=4\pi F({\bf r}),
\Delta=\left({\partial\over\partial{\bf r}} \right)^2.
\end{equation}
\section{The equation system for one degenerated energy level}

Let us consider $n$ times degenerated exciton energy level.
Without taking into account the electron-light interaction,
\begin{equation}
\label{52}\omega_\eta=\omega_0,
\end{equation}
and
\begin{equation}
\label{53}\gamma_\eta=\gamma, F_\eta({\bf r})=F({\bf r}),
P_\eta({\bf \kappa})=P({\bf \kappa}), \Phi_\eta({\bf r})=\Phi({\bf
r}).
\end{equation}
Only the vectors ${\bf p}_{cv\eta}$ depend on index $\eta$, taking
$n$ values. Taking into account only this level (assuming, that
the frequency of stimulating light is in the resonance with
 $\omega_0$),
we obtain from (44) and (50) 2n algebraical equations for 2n
quantities $M_\eta$ and ${\tilde M}_\eta$.
\begin{eqnarray}
\label{54}&&M_{\eta^\prime}=2\pi E_0D_0(\omega)({\bf l}_\ell{\bf
p}_{cv\eta^\prime})P^*({\bf \kappa}_\ell)\nonumber\\
&&+\sum_\eta\left\{{M_\eta\Omega({\bf p}_{cv\eta}^*,{\bf
p}_{cv\eta^\prime})\over \omega-\omega_0+i\gamma/2} +{{\tilde
M}_\eta\Omega({\bf p}_{cv\eta},{\bf
p}_{cv\eta^\prime})\over \omega+\omega_0+i\gamma/2}\right\},\nonumber\\
&&{\tilde M}_{\eta^\prime}=2\pi E_0D_0(\omega)({\bf e}_\ell{\bf
p}_{cv\eta^\prime}^*)P^*({\bf \kappa}_\ell)\nonumber\\
&&+\sum_\eta\left\{{M_\eta\Omega({\bf p}_{cv\eta}^*,{\bf
p}_{cv\eta^\prime})\over \omega-\omega_0+i\gamma/2} +{{\tilde
M}_\eta\Omega({\bf p}_{cv\eta},{\bf p}_{cv\eta^\prime})\over
\omega+\omega_0+i\gamma/2}\right\},~~~~
\end{eqnarray}
where
\begin{equation}
\label{55}\Omega({\bf p}_1,{\bf p}_2)=\Omega({\bf p}_2,{\bf
p}_1)=\Delta\omega({\bf p}_1,{\bf p}_2)-i\gamma_r({\bf p}_1,{\bf
p}_2)/2,
\end{equation}
\begin{eqnarray}
\label{56}\Delta\omega({\bf p}_1,{\bf p}_2)={e^2\over 2\pi^2\hbar
c}{\omega\over\omega_gm_0^2c\kappa^2}\int d^3k |P({\bf
k})|^2\nonumber\\
\times\left\{({\bf p}_1,{\bf p}_2)+[({\bf k}_1{\bf p}_1)({\bf
k}{\bf p}_2)-k^2({\bf p}_1,{\bf p}_2)]{P\over \kappa^2-k^2}
\right\},
\end{eqnarray}
\begin{eqnarray}
\label{57}\gamma_r({\bf p}_1,{\bf p}_2)={e^2\over 2\pi\hbar
c}{\omega\over\omega_gm_0^2c\kappa}\int do_{{\bf \kappa}} |P({\bf
k})|^2\nonumber\\
\times\{\kappa^2({\bf p}_1{\bf p}_2)-({\bf \kappa}{\bf p}_1)({\bf
\kappa}{\bf p}_2) \}.
\end{eqnarray}
\section{The exciton $\Gamma_6\times\Gamma_7$ in cubic crystals of $T_d$ class}

As an example, let us consider an exciton, formed by an electron
from twice degenerated $ \Gamma_6 $ conduction band and by a hole
from twice degenerated $ \Gamma_7 $ valence band, splitted by the
spin - orbital interaction.

The electron wave functions have the structure \cite{14}
\begin{equation}
\label{58}\Psi_{c1}=iS\uparrow,  \Psi_{c2}=iS\downarrow,
\end{equation}
and the the hole wave functions are
\begin{eqnarray}
\label{59}\Psi_{h1}={1\over\sqrt{3}}(X-iY)\uparrow-{1\over\sqrt{3}}Z\downarrow,\nonumber\\
\Psi_{h2}={1\over\sqrt{3}}(X+iY)\downarrow+{1\over\sqrt{3}}Z\uparrow.
\end{eqnarray}
Combining  (58) и (59) in pairs, we obtain four times degenerated
excitonic state, for which the vectors ${\bf p}_{cv\eta}$ are
\begin{eqnarray}
\label{60}&&{\bf p}_{cv1}={p_{cv}\over\sqrt{3}}({\bf e}_x-i{\bf
e}_y),\nonumber\\
&&{\bf p}_{cv2}={p_{cv}\over\sqrt{3}}({\bf e}_x+i{\bf
e}_y),\nonumber\\
&&{\bf p}_{cv3}={p_{cv}\over\sqrt{3}}{\bf e}_z,\nonumber\\
&&{\bf p}_{cv4}=-{p_{cv}\over\sqrt{3}}{\bf e}_z,
\end{eqnarray}
where we introduced the scalar
\begin{eqnarray}
\label{61}p_{cv}=i\langle S|{\hat p}_x|X\rangle ,
\end{eqnarray}
${\bf e}_x, {\bf e}_y, {\bf e}_z$ are the unite vectors along the
crystallographic axes.

We consider the circular polarization of stimulating and scattered
light, i.e.,
\begin{eqnarray}
\label{62}{\bf e}_\ell^{\pm}={1\over\sqrt{2}}({\bf e}_{xl}\pm{\bf
e}_{yl} ),\nonumber\\
{\bf e}_s^{\pm}={1\over\sqrt{2}}({\bf e}_{xs}\pm{\bf e}_{ys} ),
\end{eqnarray}
where the unite vectors ${\bf e}_{xl}$  and  ${\bf e}_{yl}$ are
perpendicular to the axis $z_\ell$ along the vector ${\bf
\kappa}_{l}$ , the unite vectors ${\bf e}_{xs}$ and ${\bf e}_{ys}$
are perpendicular to the axis $z_s$ along the vector ${\bf
\kappa}_{s}$.

In the case of the exciton $\Gamma_6\times\Gamma_7$, it is
conveniently  to sum in (3) and (6) on indexes $\eta$, using (55).
Calculations, described in sections  III - XIII , result in the
same expressions, certainly. After summation, we have
\begin{eqnarray}
\label{63}&&{\bf j}({\bf r}, t)={ie^2p_{cv}^2\over
3\pi\hbar\omega_gm_0^2}F({\bf r})\int_{-\infty}^\infty d\omega
e^{-i\omega t}{\bf T}(\omega)L(\omega)\nonumber\\&&+c.c.,
\end{eqnarray}
\begin{eqnarray}
\label{64}&&\rho({\bf r}, t)={ie^2p_{cv}^2\over
3\pi\hbar\omega_gm_0^2}F({\bf r})\int_{-\infty}^\infty
{d\omega\over\omega} \left({dF({\bf r})\over d{\bf r}} \right){\bf
T}(\omega)L(\omega)\nonumber\\&&+c.c.,
\end{eqnarray}
where the designations are introduced
\begin{eqnarray}
\label{65}&&{\bf T}(\omega)=\int d^3r{\bf {\cal E}}({\bf r},
\omega)F({\bf
r}),\nonumber\\
&&L(\omega)={1\over \omega-\omega_0+i\gamma/2}+{1\over
\omega+\omega_0+i\gamma/2}.
\end{eqnarray}
Expressions (63) and (64) do not contain evidently the vectors
${\bf r}_{cv\eta}$ and allow to calculate the light scattering
cross section by a quantum dot in the resonance with the exciton
$\Gamma_6\times\Gamma_7$. It follows from (63) and (64), that in
that case all the values do not depend on the crystallographic
axes direction.

Using (7) and (8), we obtain vector and scalar potentials
\begin{eqnarray}
\label{66}&&{\bf A}({\bf r}, t)={ie^2p_{cv}^2\over
3\pi\hbar\omega_gm_0^2c}\int_{-\infty}^\infty d\omega e^{-i\omega
t}L(\omega){\bf T}(\omega)\nonumber\\
&&\times\int d^3r^\prime F({\bf r}^\prime){e^{i\kappa|{\bf r}-{\bf
r}^\prime|}\over |{\bf r}-{\bf r}^\prime|}+c.c.,
\end{eqnarray}
\begin{eqnarray}
\label{67}&&\varphi({\bf r}, t)={ie^2p_{cv}^2\over
3\pi\hbar\omega_gm_0^2\nu^2}\int_{-\infty}^\infty
{d\omega\over\omega} e^{-i\omega t}L(\omega){\bf T}(\omega)
\nonumber\\
&&\times{d\over d{\bf r}}\int d^3r^\prime F({\bf
r}^\prime){e^{i\kappa|{\bf r}-{\bf r}^\prime|}\over |{\bf r}-{\bf
r}^\prime|}+c.c..
\end{eqnarray}
For electric and magnetic fields, one can apply (19) - (22) with
the substitution
\begin{equation}
\label{68}Q({\bf \kappa}_s, \omega, {\bf e}_s^\pm)={2p_{cv}^2\over
3}P({\bf \kappa}_s)({\bf T}(\omega){\bf e}_s)L(\omega).
\end{equation}
For the Pointing vector at $r\rightarrow\infty$, one has to use
 (27) with substitution (68). Для сечения
рассеяния при монохроматическом облучении применимы точные формулы
Formulas (33) - (34) with substitution (68) are applicable for the
scattering section at a monochromatic irradiation. In the lowest
order on the electron-light interaction, we obtain
\begin{eqnarray}
\label{69}&&{d\sigma_\mu\over do_s}={4\over 9}\left({e^2\over\hbar
c} \right)^2{p_{cv}^4\omega_\ell^2\over\omega_g^2m_0^4c^2}|{\bf
e}_\ell){\bf e}_s^*|^2|P({\bf \kappa}_\ell)|^2|P({\bf \kappa}_s)|^2\nonumber\\
&&\times \left|{1\over \omega-\omega_0+i\gamma/2}+{1\over
\omega+\omega_0+i\gamma/2}\right|^2,
\end{eqnarray}
and
$$|{\bf
e}_\ell^+{\bf e}_s^-|^2={1\over
4}(1+\cos\theta)^2,~~~~~~~~~~~~~~~~~~~~~~~~~$$
$$|{\bf
e}_\ell^+{\bf e}_s^+|^2={1\over
4}(1-\cos\theta)^2,~~~~~~~~~~~~~~~~~~~~~(69a)$$
where  $\theta$  is the scattering angle, what coincide with
results (32) - (35) of the article \cite{bb14}, if in(69) the
nonresonant contribution (proportional to
$(\omega+\omega_0+i\gamma/2)^{-1}$) is omitted, and the factor
$(\omega_\ell/\omega_g)^2$  is substituted by
$(\omega_\ell/\omega_g)^4$, and $(\omega-\omega_0+i\gamma/2)^{-1}$
is substituted by $(\omega+\omega_0+i\delta)^{-1},
\delta\rightarrow 0$ .

\section{The equation for the electric field inside of a quantum dot
in the case of the exciton $\Gamma_6\times\Gamma_7$}

Having substituted expressions (29), (66) and (67) in (41), we
obtain the equation for the Fourier-component of the electric
field

Подставив в (41) выражения (29), (66) и (67), получим уравнение
для Фурье-компонент     электрического поля
\begin{eqnarray}
\label{70}&&{\bf {\cal E}}({\bf r}, \omega)=2\pi E_0{\bf e}_\ell
e^{i{\bf \kappa}_\ell{\bf r}}D_0(\omega)-{2\over 3}{e^2\over \hbar
c}{p_{cv}^2\omega\over m_0^2\omega_g c}L(\omega)
\nonumber\\
&&\times \left[{\bf T}(\omega)+\kappa^{-2}\left({d\over d{\bf
r}}{\bf T}(\omega)\right) \right]\Phi({\bf r}).
\end{eqnarray}
The equation (70) can be obtained also from (42) with the help of
substitution the vector $ {\bf p} _ {cv\eta} $ from (60) and
summation on indexes $ \eta $, taking the values from 1 to 4.

Let us multiply (70)on $F({\bf r})$ and integrate on ${\bf r}$. We
obtain the equation for the vector ${\bf T}(\omega)$
\begin{eqnarray}
\label{71}&&{\bf T}(\omega)=2\pi E_0{\bf e}_\ell D_0P^*({\bf
r}_\ell)
\nonumber\\
&&-C(\omega)\int d^3r \left[{\bf T}(\omega)\Phi({\bf
r})+\kappa^{-2}\left({d\over d{\bf r}}{\bf T}(\omega)\right)
\Phi({\bf r})\right],~~~~~~
\end{eqnarray}
where the designation
\begin{equation}
\label{72}C(\omega)={2\over 3}\left({e^2\over\hbar c}
\right){p_{cv}^2\omega\over\omega_g^2m_0^2c}L(\omega),
\end{equation}
is introduced, and expression (50) is used for the function
$\Phi({\bf r})$. Substituting (50) in (71), we obtain
\begin{eqnarray}
\label{73}&&{\bf T}(\omega)(1+C(\omega))\int d^3k J({\bf k})= 2\pi
E_0{\bf e}_\ell D_0P^*({\bf \kappa}_\ell)
\nonumber\\
&&+{C(\omega)\over\kappa^2}\int d^3k{\bf k}({\bf k}{\bf
T}(\omega))J({\bf k}),
\end{eqnarray}
\begin{equation}
\label{74} J({\bf k})=|P({\bf k})|^2 \left[{1\over 2\pi^2}{P\over
k^2-\kappa^2}+{i\over 4\pi\kappa}\delta(k-\kappa) \right].
\end{equation}
Expression (73) may be treated as a system of three equations for
three components $ T_x(\omega), T_y(\omega), T_z(\omega)$ of
vector
 ${\bf T}(\omega)$. The result must be substituted into the formula for the scattering section
 at monochromatic excitation or into the formula for the Pointing vector at $r\rightarrow\infty$
 at pulse excitation.
Thus, thus the problem of the resonant light scattering is solved.

\section{Light scattering by the exciton
 $ \Gamma_6\times\Gamma_7 $ in a spherical quantum dot}

Solution of the problem is simplified, when $P({\bf k})$ depends
only on the magnitude of the vector ${\bf k}$ in the case of a
spherically symmetrical function
\begin{equation}
\label{75}F({\bf r})=F(r).
\end{equation}
Then, at $J({\bf k})=J(k)$,
\begin{eqnarray}
\label{76}\int d^3kk_xk_y J(k)=\int d^3kkk_xk_z
J(k)\nonumber\\=\int d^3k kk_yk_zJ({\bf k})=0,
\end{eqnarray}
and the solution of (73) has the form
\begin{eqnarray}
\label{77}&&{\bf T}(\omega)=\nonumber\\
&&={2\pi E_0{\bf e}_\ell D_0(\omega)P^*( \kappa)\over
1-(\Delta\omega_0-i\gamma_r/2)\left[{1\over\omega-\omega_0+i\gamma/2}+{1\over\omega-\omega_0+i\gamma/2}\right
]},~~~~
\end{eqnarray}
where
\begin{eqnarray}
\label{78}&&\Delta\omega_0=\Delta\omega_0^\prime+\Delta\omega_0^{\prime\prime},\nonumber\\
&&\Delta\omega_0^{\prime}
 = {4\over
9\pi}{e^2p_{cv}^2\over\hbar\omega_gm_0^2\nu^2}\int_0^\infty dk
k^3|P(k)|^2{P\over\kappa-k},
\end{eqnarray}
\begin{equation}
\label{79}\Delta\omega_0^{\prime\prime}
 = {4\over
9\pi}{e^2p_{cv}^2\over\hbar\omega_g\omega m_0^2\nu^2}\int_0^\infty
dk k^2|P(k)|^2\left(3-{k\over\kappa+k}\right),
\end{equation}
\begin{equation}
\label{80}\gamma_r
 = {8\over
9}{e^2p_{cv}^2\omega^2\nu\over\hbar\omega_gm_0^2c^3}|P(\kappa)|.
\end{equation}
If the nonresonant contribution $(\omega+\omega_0+i\gamma/2)^{-1}$
in (77) is omitted, we obtain
\begin{equation}
\label{81}{\bf T}(\omega)={2\pi E_0{\bf e}_\ell
D_0(\omega)P^*(\kappa)(\omega-\omega_0+i\gamma/2)\over
\omega-(\omega_0+\Delta\omega_0)+i(\gamma+\gamma_r)/2}.
\end{equation}
Omitting nonresonant contributions and using (81), we obtain the
results (19), (20) and (22) for electric and magnetic fields at
$\rightarrow\infty$ with the help of substitution
\begin{eqnarray}
\label{82}&&E^\pm({\bf r}, t)=-{2\over
3}{E_0e^2p_{cv}^2\omega^2\nu\over\hbar\omega_gm_0^2c^2r}({\bf
e}_\ell{\bf e}_s^\pm)\nonumber\\&&\times\int_{-\infty}^\infty
d\omega{\omega e^{i\kappa r-i\omega
t}D_0(\omega)|P(\kappa)|^2\over
\omega-(\omega_0+\Delta\omega_0)+i(\gamma+\gamma_r)/2}.
\end{eqnarray}
For the Pointing vector at $r\rightarrow\infty$ we obtain
\begin{eqnarray}
\label{83}&&{\bf S}({\bf r}, t, {\bf e}_s)={2\over
9\pi}E_0^2\left({e^2\over\hbar c}\right)^2
{p_{cv}^4\nu\over\omega_g^2m_0^4c}|{\bf e}_\ell{\bf
e}_s^\pm|^2{{\bf
r}\over r^3} \nonumber\\
 &&\times\left|\int_{-\infty}^\infty d\omega{\omega e^{i\kappa
r-i\omega t}D_0(\omega)|P(\kappa)|^2\over
\omega-(\omega_0+\Delta\omega_0)+i(\gamma+\gamma_r)/2}\right|^2.
\end{eqnarray}
At last, the result for the scattering section at monochromatic
irradiation is as follows
\begin{eqnarray}
\label{84}&&{d\sigma_\mu\over do_s}={4\over 9}\left({e^2\over\hbar
c}\right)^2 {p_{cv}^4\omega_\ell^2\over\omega_g^2m_0^4c^2}|{\bf
e}_\ell{\bf e}_s^\pm|^2 \nonumber\\
 &&\times{|P(\kappa)|^4\over
[\omega_\ell-(\omega_0+\Delta\omega_0)]^2+(\gamma+\gamma_r)^2/4}..
\end{eqnarray}
Thus, the precise results (82) - (84) distinguish from the results
of the lowest approximation on the electron-light interaction only
by the replacement of  $\omega_0$ на $\omega_0+\Delta\omega$ и
$\gamma$ by $\gamma+\gamma_r$ .

\section{Comparison of results for radiative damping and amendments
 to the energy with results of the quantum perturbation theory}

 Let us compare expression (80) for $\gamma_r$ with the expression
(43) from \cite{bb5}. They coincide almost. Thus, one can conclude
that $\gamma_r$ from (82) - (84) coincides with $\gamma_{r\eta}$
with indices  $\eta=1$ or $\eta=2$ from \cite{bb5}.

As far as concerns energy amendments, we have no a coincidence
with the results of the quantum theory. The calculation shows that
the energy amendment of the perturbation theory, taking into
account the exciton - photon transition, must be equal
\begin{eqnarray}
\label{85}&&\Delta\omega_\eta={e^2\over 4\pi^2\hbar
\omega_g^2m_0^2\nu^2}\int do_{{\bf k}}\sum_\mu
|{\bf p}_{cv\eta}{\bf e}_{\mu{\bf k}}|^2 \nonumber\\
 &&\times\int dk k^3|P_\eta({\bf k})|^2{P\over k_\eta-k},
\end{eqnarray}
where $k_\eta=\omega_\eta\nu/c$.

For the exciton  $\Gamma_6\times\Gamma_7$   we use vectors ${\bf
p}_{cv\eta}$ , defined in (60). Assuming  $P_\eta({\bf k})=P({\bf
k})=P( k) $ , we have
\begin{eqnarray}
\label{86}&&\Delta\omega_1=\Delta\omega_2={4\over
9\pi}{e^2p_{cv}^2\over \hbar\omega_g^2m_0^2\nu^2}\int dk
k^3|P_\eta({\bf k})|^2{P\over k_\eta-k},\nonumber\\
&&\Delta\omega_3=\Delta\omega_4=\Delta\omega_1/2.
\end{eqnarray}
Comparing (86) and (78) and assuming $\omega_\eta=\omega$, we find
that the RHS of (86) contains an additional factor
$\omega/\omega_g$ in comparison to the RHS of (78), what is
incidentally. However, the energy amendment to
$\Delta\omega_0^{\prime\prime}$, determined in (79), is not in
agreement with the results of the perturbation theory.

\section{Results for scattering sections}

For the scattering section on excitons $\Gamma_6\times\Gamma_7$ in
spherically symmetrical quantum dots, we obtain with the help of
(84) and (69a)
\begin{eqnarray}
\label{87}&&{d\sigma^{++}\over do_s}={d\sigma^{++}\over
do_s}={1\over 9}{\tilde\Sigma}_0(1+\cos\theta)^2,\nonumber\\
&&{d\sigma^{+-}\over do_s}={d\sigma^{-+}\over
do_s}={1\over 9}{\tilde\Sigma}_0(1-\cos\theta)^2,\nonumber\\
&&{\tilde\Sigma}_0=\left({e^2\over\hbar c}
\right)^2\left({\omega_\ell\over\omega_g} \right)^2 {p_{cv}^4\over
m_0^4c^2}{|P(k_\ell)|\over
(\omega_\ell-{\tilde\omega}_0)^2+\Gamma^2/4},
\nonumber\\
&&{\tilde\omega}_0=\omega_0+\Delta\omega,  \Gamma=\gamma+\gamma_r,
k_\ell={\omega_\ell\nu\over c},
\end{eqnarray}
the superscript $++$  designates the polarization of incident
 (scattered) light ${\bf e}^+_\ell({\bf
e}^+_s)$ and so on.

Summing on polarizations of scattered light, we obtain
\begin{equation}
\label{88}{d\sigma^{+}\over do_s}={d\sigma^{-}\over do_s}={2\over
9}{\tilde\Sigma}_0(1+\cos^2\theta),
\end{equation}
where superscript $+(-)$  designates  the polarization of incident
 light
${\bf e}^+_\ell({\bf e}^+_\ell)$ .

The total scattering section is equal равно
\begin{equation}
\label{89}\sigma^{+}=\sigma^{-}={32\pi\over 27}{\tilde\Sigma}_0.
\end{equation}
Using (80) we obtain for the radiative damping that in the
resonance at  $\omega_\ell={\tilde\omega}_0$ and under condition
$\gamma\ll\gamma_r$
\begin{equation}
\label{90}\sigma^{+}_{res}=\sigma^{-}_{res}=6\pi\left({\lambda_\ell\over
2\pi}\right)^2,
\end{equation}
where  $\lambda_\ell$  is the light wave length.The result
 (90) is correct for any sizes of a quantum dot.

For example, using the  "envelope"  wave function
\begin{equation}
\label{91}F({\bf r})={1\over 2\pi R}{\sin^2(\pi r/R)\over
r^2}\theta(R-r),
\end{equation}
corresponding to the low exciton energy level in a spherical
quantum dot with infinitely rectangular barriers, we have
\begin{equation}
\label{92}P(k)={2\over kR}\int_0^\pi
d\kappa\sin{kR_x\over\pi}{\sin^2\kappa\over \kappa}, P(0)=1.
\end{equation}
\section{Conclusion}

The problem   of elastic  light scattering by semiconductor
quantum dots of any form, sizes and configuration is solved with
the help of the semiclassical method of retarded potentials.
Results are applicable for monochromatic and pulse excitations.
The light refraction coefficients are assumed identical inside and
outside of a quantum dot.

The  electron - light interaction is taken precisely into account,
i.e., all the light reradiation and absorption processes.

As an example, the differential cross section of light scattering
is calculated for monochromatic light with frequency $ \omega_\ell
$ on spherical quantum dot in a semiconductor of $T_d $ class in
 resonance of stimulating light with exciton $
\Gamma_6\times\Gamma_7 $. It is shown that at least in case of
this example the exact account of  electron - light interaction
results only in replacement of the factor $ (\omega_l-\omega_0) ^2
+\gamma^2/4 $, obtained in lowest approximation, by a factor
 $(\omega_l-{\tilde\omega}_0)^2+(\gamma+\gamma_r)^2/4$, where
 ${\tilde\omega}_0=\Delta\omega_0+\Delta\omega$ is the exciton energy
renormalized by long-distance  by exchange interaction. The value
$ \gamma_r $ is coordinated to result received with By the help of
the quantum perturbation theory\cite{bb5}.

Under condition $k_lR\ll 1 $  ($k_l $ is the module of the light
wave vector, $R $ is the size of a quantum dot), polarization and
angular distribution of light do not depend neither on the form of
quantum dot, nor from "envelope" $F ({\bf r}) $ exciton wave
function, but only from vectors $ {\bf p} _ {cv\eta} $ of
nondiagonal matrix elements of exciton momentum of excitonic state
with an index $ \eta $, and magnitude of scattering cross section
does not depend on the sizes of quantum dot.

The results of the present work can be used for precise
calculations of light absorbance by any quantum dot, which is
proportional to nonradiative damping $ \gamma $ (see, for example,
\cite{bb8,bb9,bb10,bb12}).

At last, obtained results for the Pointing vector on large
distances from a quantum dot are applicable for the pulse
irradiations. The pulse form is determined by function $D_0
(\omega) $ (see(31)). It allows, for example, to describe
oscillations of scattered light caused by splitting  of exciton
energy levels in quantum dots (compare \cite{bb15}, where similar
oscillations are predicted in light reflection and absorption by a
quantum well at pulse irradiation).

\end{document}